\documentstyle[12pt]{article} 
\begin{document} 
{\pagestyle{empty} 


\vskip 25mm 

\centerline{\large \bf Star-Triangle Relation}
\centerline{\large \bf of the Chiral Potts Model Revisited}

\vskip 10mm

\vskip 20mm

\centerline{Minoru Horibe \footnote{E-mail address:
horibe@edu00.fukui-u.ac.jp}}
\centerline {{\it Department of Applied Physics, 
Faculty of Engineering}}
\centerline {{\it Fukui University, Fukui 910, Japan}}

\vskip 3mm

\centerline{Kazuyasu Shigemoto \footnote{E-mail address:
shigemot@tezukayama-u.ac.jp}}
\centerline {{\it Department of Physics}}
\centerline {{\it Tezukayama University, Nara 631, Japan }}

\vskip 2cm

\centerline{\bf Abstract} 

\vskip 10mm

We give the simple proof of the star-triangle relation of the 
chiral Potts model. We also give the constructive way to understand 
the star-triangle relation of the chiral Potts model, which may 
give the hint to give the new integrable models. 

\newpage}


\section{Introduction}

\indent

The Ising model is the most important integrable
model in two-dimension\cite{Onsager}, but this model 
has various nice properties so
that we cannot guess what kind of properties are inherited in the more
general two-dimensional integrable model. In this sense, the 
eight-vertex model\cite{Baxter1} and the chiral Potts 
model\cite{McCoy,Baxter2}, which include the Ising model 
as a special case, are key models to 
understand the mechanism of the integrability in two-dimension.
Moreover, the methods for analyzing these models are applicable
to another ones and we can obtain various integrable models in 
two-dimension by modifying these key models.

Epecially, the chiral Potts model is related not only to 
the two-dimensional integrable model but also to the 
three-dimensional integrable models. The Bazhanov-Baxter 
model\cite{B-B}, which is the genelalization of the 
three-dimensional integrable Zamolodohikov 
model\cite{Zamolodchikov}, is constructed 
through the two dimensional integrable $sl_n$ chiral 
Potts model\cite{B-K-M-S,D-J-M-M}. 
To understand the mathematical structure of the three 
dimensional integrable model \cite{Kashaev,ours}, we 
must clarify more about the integrability, the 
star-triangle relation, in the chiral Potts model. 
Au-Yang {\it et. al.} have given the proof of the 
star-triangle relation in the chiral Potts model\cite{AuYang}, 
but they used the complicated recursion relation so that the 
origin of the integrability is not clear in thier proof. 

In this paper, we give the simple proof of the 
star-triangle relation of the chiral Potts model.
Our method is more clear about the mathematical structure
and gives the hint to generalize it. We also give the 
constructive way of understanding the star-triangle relation 
of the chiral Potts model, which may give the new integrable 
models.



\section{Rewriting the star-triangle relation of the 
chiral Potts model}

\indent

In order to prove the star-triangle relation,
we first rewrite the star-triangle relation 
in a more convenient way to prove it in this section.
We also show various formulae, which is used in the proof.

\subsection{Preparation}

\indent

The Boltzmann weight of  $N$-state chiral Potts model is 
given by $W_{pq} (n)$ or $\overline{W}_{pq}(n)$ depending on the 
direction. Here $\left\{ p, q \right\}$ represents the rapidity, 
and $n$ is the spin variable with $n=1,2, \cdots , N$.
The Boltzmann weights are periodic with spin, that is, 
$W_{pq} (n+N)=W_{pq} (n)$ and 
$\overline{W}_{pq}(n+N)=\overline{W}_{pq}(n)$.
The star-triangle relation of the chiral Potts model 
is 
\begin{eqnarray}
 && \sum^{N-1}_{d=0} \overline{W}_{qr}(b-d) W_{pr}(a-d) 
 \overline{W}_{pq}(d-c)
 =R_{pqr} W_{pq}(a-b) \overline{W}_{pr}(b-c) W_{qr}(a-c) .
\label{e1} 
\end{eqnarray}
The Boltzmann weights $W_{pq}(n)$ and $\overline{W}_{pq}(n)$ 
in Eq.(\ref{e1}) are 
\begin{eqnarray}
 &&W_{pq}(n)
  = \rho_1 \prod^{n}_{j=1} \frac{d_p b_q -a_p c_q \omega^{j}}
 {b_p d_q -c_p a_q \omega^{j}},  \qquad
  \overline{W}_{pq}(n)
   = \rho_2 \prod^{n}_{j=1} \frac{\omega a_p d_q -d_p a_q \omega^{j}}
               {c_p b_q -b_p c_q \omega^{j}},  
\label{e2}
\end{eqnarray}
where 
\begin{eqnarray}
  &&{a_p}^N+k' {b_p}^N=k {d_p}^N,  \qquad
   k' {a_p}^N+{b_p}^N=k {c_p}^N,
\label{e3}
\end{eqnarray}
with $\omega =e^{2\pi i /N}$ and $k^2+{k'}^2=1$.

By making the recursion relation, we have the following Fourier transforms 
of $W$ and $\overline{W}$ as $\widetilde{W}$ and 
$\widetilde{\overline{W}}$ in the form 
\begin{eqnarray}
   &&\widetilde{W}_{pq}(m)
   =\sum^{N-1}_{k=0} \omega^{mk} W_{pq}(k)
   =\rho'_1 \prod^{m}_{j=1} \frac{b_p d_q -d_p b_q \omega^{j-1}}
    {c_p a_q -a_p c_q \omega^{j}},
\label{e4}\\
  &&\widetilde{\overline{W}}_{pq}(m)
   =\sum^{N-1}_{k=0} \omega^{mk} \overline{W}_{pq}(k)
   = \rho'_2 \prod^{m}_{j=1} \frac{c_p b_q -a_p d_q \omega^{j}}
    {b_p c_q -d_p a_q \omega^{j}}, 
\label{e5}
\end{eqnarray}
where 
$\displaystyle{\rho'_1=\rho_1 \sum^{N-1}_{n=0} 
\prod^{n}_{j=1} \frac{d_p b_q -a_p c_q \omega^j}
{b_p d_q -c_p a_q \omega^j}}$ and 
$\displaystyle{\rho'_2=\rho_2 \sum^{N-1}_{n=0} 
\prod^{n}_{j=1} \frac{\omega a_p d_q -d_p a_q \omega^j}
{c_p b_q -b_p c_q \omega^j}}$.

The inverse Fourier transform is 
\begin{eqnarray}
  W_{pq}(m)
   = \frac{1}{N} \sum^{N-1}_{k=0} \omega^{-mk} \widetilde{W}_{pq}(k), 
   \quad   \overline{W}_{pq}(m)
   = \frac{1}{N} \sum^{N-1}_{k=0} \omega^{-mk} 
     \widetilde{\overline{W}}_{pq}(k).   
\label{e6}
\end{eqnarray}
Using the above Fourier transform, $R_{pqr}$ is expressed 
by\cite{Baxter1,AuYang,Matveev}

\begin{eqnarray}
  R_{pqr}=f_{pq} f_{qr}/f_{pr}, \quad 
  f_{pq}
  =\left[ \prod^{N-1}_{m=0} \widetilde{\overline{W}}_{pq}(m)
        /W_{pq}(m) \right]^{1/N}. 
\label{e7} 
\end{eqnarray}
We define 
\begin{eqnarray}
  f_{pq}=f^{(1)}_{pq}/f^{(2)}_{pq} , \quad
  f^{(1)}_{pq}
  =\left[\prod^{N-1}_{m=0} 
    \widetilde{\overline{W}}_{pq}(m) \right]^{1/N}, 
  \quad 
  f^{(2)}_{pq}
  =\left[ \prod^{N-1}_{m=0} W_{pq}(m) \right]^{1/N} , 
\label{e8}
\end{eqnarray}
which gives $R_{pqr}=f^{(1)}_{pq} f^{(1)}_{qr} f^{(2)}_{pr}
    /f^{(1)}_{pr} f^{(2)}_{pq} f^{(2)}_{qr}$.
Replacing $b \rightarrow b+a$, $c \rightarrow c+a$ and 
$d \rightarrow d+a$ in Eq.(\ref{e1}), we have 
\begin{eqnarray}
 && \sum^{N-1}_{d=0} \overline{W}_{qr}(b-d) W_{pr}(-d) 
 \overline{W}_{pq}(d-c)
 =R_{pqr} W_{pq}(-b) \overline{W}_{pr}(b-c) W_{qr}(-c) .
\label{e9} 
\end{eqnarray}
We give the proof of this star-triangle relation in the 
next chapter.

\subsection{Rewriting the star-triangle relation with the cyclic 
representation}

\indent

Using the cyclic representation of $su(2)$, we rewrite the 
above expression.
The basis of the cyclic representation of $su(2)$ are $X$ and $Z$
with the properties
\begin{eqnarray}
 && ZX=\omega XZ , \quad X^N=Z^N=1 .  
\label{e10} 
\end{eqnarray}
The explicit $N \times N$ matrix representation is 
\begin{eqnarray}
 X_{\alpha, \beta}=\delta_{\alpha, \beta+1}
   +\delta_{\alpha, \beta+1-N},  \qquad
 Z_{\alpha, \beta}=\omega^{\alpha} \delta_{\alpha, \beta} .
\label{e11}
\end{eqnarray}
We define
\begin{eqnarray}
   T_{pq}=\sum^{N-1}_{m=0} \widetilde{W}_{pq}(m) Z^m, \quad
   S_{pq}=\sum^{N-1}_{m=0} \overline{W}_{pq}(m) X^m,
\label{e12}
\end{eqnarray}
and make the following quantities
\begin{eqnarray}
&& T_{pq} S_{pr} T_{qr}=\sum_{l,m,n} \widetilde{W}_{pq}(l)
   \overline{W}_{pr}(m) \widetilde{W}_{qr}(n) 
   \omega^{lm} X^m Z^{l+n},   
\label{e13}\\
&& S_{qr} T_{pr} S_{pq}=\sum_{l,m,n} \overline{W}_{qr}(l)
   \widetilde{W}_{pr}(m) \overline{W}_{pq}(n) 
   \omega^{mn} X^{l+n} Z^m .   
\label{e14}  
\end{eqnarray}
We take the component of these operators and obtain 
\begin{eqnarray}
  \left(T_{pq} S_{pr} T_{qr} \right)_{bc}
  &=&\sum_{l,m,n} \sum_{d} \widetilde{W}_{pq}(l)
   \overline{W}_{pr}(m) \widetilde{W}_{qr}(n) 
   \omega^{lm} (X^m)_{bd} (Z^{l+n})_{dc}   \nonumber\\
  &=&\sum_{l,n} \widetilde{W}_{pq}(l)  
   \overline{W}_{pr}(b-c) \widetilde{W}_{qr}(n) 
   \omega^{lb+cn} \nonumber\\
  &=&N^2 W_{pq}(-b) \overline{W}_{pr}(b-c) 
     W_{qr}(-c)  ,  
\label{e15}\\
   \left(S_{qr} T_{pr} S_{pq}\right)_{bc}
   &=&\sum_{l,m,n} \sum_{d} \overline{W}_{qr}(l)
   \widetilde{W}_{pr}(m) \overline{W}_{pq}(n) 
   \omega^{mn} (X^{l+n})_{bd}  (Z^m)_{dc}  \nonumber\\
  &=& \sum_{m,n} \overline{W}_{qr}(b-c-n)
   \widetilde{W}_{pr}(m) \overline{W}_{pq}(n) 
   \omega^{m(n+c)}    \nonumber\\
  &=& N \sum_{n} \overline{W}_{qr}(b-c-n)
   W_{pr}(-n-c) \overline{W}_{pq}(n)   \nonumber\\
  &=& N \sum_{d} \overline{W}_{qr}(b-d)
   W_{pr}(-d) \overline{W}_{pq}(d-c) .
\label{e16}  
\end{eqnarray}
Then the star-triangle relation is rewritten as 
\begin{eqnarray}
  && S_{qr} T_{pr} S_{pq}=R_{pqr} T_{pq} S_{pr} T_{qr}/N , 
\label{e17}
\end{eqnarray}
which becomes
\begin{eqnarray}
  && N S_{qr} T_{pr} S_{pq}/f^{(1)}_{qr} f^{(2)}_{pr} f^{(1)}_{pq}
  =T_{pq} S_{pr} T_{qr}/f^{(2)}_{pq} f^{(1)}_{pr} f^{(2)}_{qr} , 
\label{e18}
\end{eqnarray}
by using Eq.(\ref{e7}) and Eq.(\ref{e8}).
Then if we define the following quantities $\widehat{T}_{pq}$
and  $\widehat{S}_{pq}$ 
\begin{eqnarray}
  && \widehat{T}_{pq}
  =\frac{T_{pq}}{N \left( \prod^{N-1}_{m=0}W_{pq}(m)\right)^{1/N}} 
  =\frac{T_{pq}}{N f^{(2)}_{pq}} , 
\label{e19}\\
  && \widehat{S}_{pq}
   =\frac{S_{pq}}{\left( \prod^{N-1}_{m=0} 
    \widetilde{ \overline{W}}_{pq}(m)\right)^{1/N}}
   =\frac{S_{pq}}{f^{(1)}_{pq}} , 
\label{e20}
\end{eqnarray}
we can rewrite the original star-triangle relation into the 
following form
\begin{eqnarray}
  && \widehat{S}_{qr} \widehat{T}_{pr} \widehat{S}_{pq}
  =\widehat{T}_{pq} \widehat{S}_{pr} \widehat{T}_{qr}.  
\label{e21}
\end{eqnarray}

\subsection{Properties of $\widehat{T}_{pq}$ 
and  $\widehat{S}_{pq}$}

\indent

Here we show various properties of these 
$\widehat{T}_{pq}$ and  $\widehat{S}_{pq}$.
Date {\it et. al.}\cite{Date} have given the similar formula
of ours through the help of the recursion relation. 
We use the explicit representation of $Z, X$, and this method 
is more apparent than that of Date {\it et. al.}.

We first rewrite $T_{pq}$ in the following form
\begin{eqnarray}
  T_{pq}
 &=& \sum^{N-1}_{l=0} \widetilde{W}_{pq}(l) Z^l
  =\left( \begin{array} {llll} 
  \sum^{N-1}_{l=0} \widetilde{W}_{pq}(l)  & 0 & 0 & \cdots \\
  0 & \sum^{N-1}_{l=0} \widetilde{W}_{pq}(l) \omega^{l} & 0 & \cdots \\
  0 & 0 & \sum^{N-1}_{l=0} \widetilde{W}_{pq}(l) \omega^{2 l} & \cdots  \\
  \cdots  & \cdots & \cdots & \cdots \end{array}\right)  \nonumber\\
    &=&N \left( \begin{array} {llll} 
  W_{pq}(0)  & 0 & 0 & \cdots \\
  0 & W_{pq}(-1)  & 0 & \cdots \\
  0 & 0 & W_{pq}(-2) & \cdots  \\
  \cdots  & \cdots & \cdots & \cdots \end{array}\right) . 
\label{e22}
\end{eqnarray}
If we use the relation $ W_{pq}(n) W_{qp}(n) ={\rho_1}^2 \ 
(n-{\rm independent})$,
we have $ T_{pq} T_{qp} =N^2 {\rho_1}^2 \times {\bf 1}$, and 
the determinant is 
\begin{eqnarray}
  \det( T_{pq}) =N^N  \prod^{N-1}_{l=0} W_{pq}(-l) 
   = N^N \prod^{N-1}_{l=0} W_{pq}(l). 
\label{e23}
\end{eqnarray}
Then we define the normalized quantity
\begin{eqnarray}
  \widehat{T}_{pq}
  =\frac{T_{pq}}{N \left| \prod^{N-1}_{m=0}W_{pq}(m)\right|^{1/N}} 
   =\frac{T_{pq}}{\left|\det(T_{pq})\right|^{1/N}} , 
\label{e24}
\end{eqnarray}
which satisfies $\det(\widehat{T}_{pq})=1$.
Using this normalized quantity, we have 
$ \widehat{T}_{pq} \widehat{T}_{qp}={\bf 1}$.

Similarly, we can prove 
$ \widehat{S}_{pq} \widehat{S}_{qp}={\bf 1}$ in the following way.
By diagonalizing $X$ by the unitary matrix $U$ in the form 
$U^{-1} X U=Z$, we have 
\begin{eqnarray}
  S_{pq}
 &=& \sum^{N-1}_{l=0} \overline{W}_{pq}(l) X^l
  = U \left( \begin{array} {llll} 
  \sum^{N-1}_{l=0} \overline{W}_{pq}(l)  & 0 & 0 & \cdots \\
  0 & \sum^{N-1}_{l=0} \overline{W}_{pq}(l) \omega^{l} & 0 & \cdots \\
  0 & 0 & \sum^{N-1}_{l=0} \overline{W}_{pq}(l) \omega^{2 l} & \cdots  \\
  \cdots  & \cdots & \cdots & \cdots \end{array}\right) U^{-1}  \nonumber\\
    &=& U \left( \begin{array} {llll} 
 \widetilde{ \overline{W}}_{pq}(0)  & 0 & 0 & \cdots \\
  0 & \widetilde{ \overline{W}}_{pq}(-1)  & 0 & \cdots \\
  0 & 0 & \widetilde{ \overline{W}}_{pq}(-2) & \cdots  \\
  \cdots  & \cdots & \cdots & \cdots \end{array}\right) U^{-1} .
\label{e25}
\end{eqnarray}
If we use the relation 
$ \widetilde{\overline{W}}_{pq}(n) \widetilde{\overline{W}}_{qp}(n) 
={\rho'_2}^2 \ (n-{\rm independent})$,
we have $ S_{pq} S_{qp} ={\rho'_2}^2 \times {\bf 1}$, and 
the determinant is 
\begin{eqnarray}
  \det( S_{pq}) = \prod^{N-1}_{l=0} \widetilde{ \overline{W}}_{pq}(-l) 
   =  \prod^{N-1}_{l=0} \widetilde{ \overline{W}}_{pq}(l). 
\label{e26}
\end{eqnarray}
Then we define the normalized quantities
\begin{eqnarray}
  \widehat{S}_{pq}
  =\frac{S_{pq}}{\left| \prod^{N-1}_{m=0} 
  \widetilde{ \overline{W}}_{pq}(m)\right|^{1/N}}
  =\frac{ S_{pq}}{\left|\det(S_{pq})\right|^{1/N}} , 
\label{e27}
\end{eqnarray}
which satisfies $\det(\widehat{S}_{pq})=1$.
Using this normalized quantity, we have 
$ \widehat{S}_{pq} \widehat{S}_{qp}={\bf 1}$.

In this way, we summarize the properties of
$\widehat{T}_{pq}$ and $\widehat{S}_{qp}$
in the following way

\begin{eqnarray}
&& \det(\widehat{T}_{pq})=1, \quad \det(\widehat{S}_{pq})=1 , 
   \nonumber\\
&& \widehat{T}_{qp}=\widehat{T}^{-1}_{pq},  \quad
 \widehat{S}_{qp}=\widehat{S}^{-1}_{pq} .
\label{e28}
\end{eqnarray}

\subsection{Special case: the Ising model}

\indent

The Ising model is the $N=2$ case of the chiral Potts mode\cite{AuYang}. 
The star-triangle relation is \cite{Baxter3}
\begin{eqnarray}
  \sum_{d=\pm 1} \exp \{d (L_1 a+K_2 b+L_3 c) \}
  =R \exp \{K_1 b c+L_2 c a + K_3 a b \} ,
\label{e29}
\end{eqnarray}
where $a, b, c$ takes $\pm 1$. We exchange the symbol 
of $L_2$ and $K_2$ from the standard notation in order 
to compare the Ising model with the chiral Potts model.
$\{ K_1, L_2, K_3, R\}$ are expressed by $\{ L_1, K_2, L_3 \}$
in the following way
\begin{eqnarray}
 &&\exp(4 K_1)=\frac{\cosh(L_1+K_2+L_3) \cosh(-L_1+K_2+L_3)}
               {\cosh(L_1-K_2+L_3) \cosh(L_1+K_2-L_3)} , 
\label{e30}\\
 &&\exp(4 L_2)=\frac{\cosh(L_1+K_2+L_3) \cosh(L_1-K_2+L_3)}
               {\cosh(-L_1+K_2+L_3) \cosh(L_1+K_2-L_3)} , 
\label{e31}\\
 &&\exp(4 K_3)=\frac{\cosh(L_1+K_2+L_3) \cosh(L_1+K_2-L_3)}
               {\cosh(-L_1+K_2+L_3) \cosh(L_1-K_2+L_3)}  ,
\label{e32}\\
 &&R= \sqrt{2 \sinh( 2 L_1) \sinh( 2 L_3)/\sinh( 2 L_2)} ,
\label{e33}
\end{eqnarray}
where $L_2$ in Eq.(\ref{e33}) must be expressed by 
 $\{ L_1, K_2, L_3 \}$ by using Eq.(\ref{e31}).
We rewrite Eq.(\ref{e29}) by using the formula    
\begin{eqnarray}
 &&\exp( L_i a b)=\sqrt{2 \sinh(2 L_i) } 
  \left( \exp( L^{*}_i \sigma_x ) \right)_{ab} ,  
\label{e34}\\
 &&\delta_{d d'} \exp( K_i d b)
=\left( \exp( K_i b \sigma_z) \right)_{d d'} ,
\label{e35}
\end{eqnarray}
where the dual variables $L^{*}_i$ are defined by
 $\tanh {L^{*}_i}=\exp \{-2 L_i\}$. Then the star-triangle relation
is expressed by
\begin{eqnarray}
  && \sum_{d d'} \sqrt{2 \sinh(2 L_3)} 
  \left(\exp \left(L^{*}_3 \sigma_x \right) \right)_{c d} 
  \left( \exp \left( K_2 b  \sigma_z \right) \right)_{d d'}
 \sqrt{2 \sinh(2 L_1)} 
 \left(\exp \left(L^{*}_1 \sigma_x \right) \right)_{d' a} 
  \nonumber\\
 &&=R \sum_{d d'} \left( \exp \left(K_1 b \sigma_z \right) \right)_{c d} 
  \sqrt{2 \sinh(2 L_2)} 
 \left(\exp \left(L^{*}_2 \sigma_x \right) \right)_{d d'}
  \left( \exp \left(K_3 b \sigma_z \right) \right)_{d' a} .
\label{e36}
\end{eqnarray}
Using Eq.(\ref{e33}), this relation can be rewritten 
in the following form
\begin{eqnarray}
   \left( \exp \{ L_3^{*} \sigma_x \}
   \exp \{ b  K_2  \sigma_z \}
   \exp \{ L_1^{*} \sigma_x \} \right)_{c a}
  = \left( \exp \{ b  K_1  \sigma_z \}
    \exp \{ L_2^{*} \sigma_x \}
    \exp \{ b  K_3  \sigma_z \} \right)_{c a}. 
\label{e37}
\end{eqnarray}
If we notice that $b= \pm  1$, we have the star-triangle relation
in the following simple form 
\begin{eqnarray}
    \exp \{ L_3^{*} \sigma_x \}
   \exp \{ \pm K_2  \sigma_z \}
   \exp \{ L_1^{*} \sigma_x \} 
  = \exp \{ \pm  K_1  \sigma_z \}
    \exp \{ L_2^{*} \sigma_x \}
    \exp \{ \pm K_3  \sigma_z \} ,
\label{e38}
\end{eqnarray}
which is the Euclidean version of the spherical trigonometry 
relation.  

\section{The star-triangle relation in 
the chiral Potts model}

\indent

In this chaper, we first give the simple proof 
of the star-triagle relation of the chiral Potts model.
Starting from the basis $\{X, XZ, Z\}$ of the cyclic 
representation of $su(2)$, we next give the consructive proof
of the star-triangle relation and give the same expression 
of $T_{pq}$ and $S_{pq}$ as in Eq.(\ref{e24}) and 
Eq.(\ref{e27}).
We also demonstrate our method to give some generalization
of the chiral Potts model. 

\subsection{The Proof of the star-triangle relation}

\indent

From the expression of $T_{pq}$ and $S_{pq}$ in Eq.(\ref{e12}),
we can show the following formula
\begin{eqnarray}
  &&T_{pq} \left(d_p b_q Z/\omega -a_p c_q \right)X 
  =\left( b_p d_q Z/ \omega -c_p a_q \right) X  T_{pq},
\label{e39} \\
  &&S_{pq} \left(\omega a_p d_q X -c_p b_q \right) Z 
  =\left(\omega d_p a_q X -b_p c_q \right) Z S_{pq}.
\label{e40}
\end{eqnarray}
Using the above relation, we first operate  $T_{pq} S_{pr} T_{qr}$
to the following quantity $I_{p q r}$ 
\begin{eqnarray}
  I_{p q r}=-a_p a_q c_r X+a_p d_q b_r X Z  
   -c_p b_q b_r Z/ \omega .
\label{e41}
\end{eqnarray}
Then we can show 
\begin{eqnarray}
  T_{qr}I_{p q r}=I_{p r q} T_{qr}, \quad
  S_{pr}I_{p r q}=I_{r p q} S_{pr}, \quad
  T_{pq}I_{r p q}=I_{r q p} T_{pq}, 
\label{e42}
\end{eqnarray}
by using Eq.(\ref{e39}) and Eq.(\ref{e40}). This gives 
\begin{eqnarray}
 T_{pq} S_{pr} T_{qr} I_{p q r}=I_{r q p} T_{pq} S_{pr} T_{qr}.
\label{e43}
\end{eqnarray}
Similarly we operate $S_{qr} T_{pr} S_{pq}$ to the same quantity 
$I_{p q r}$ and we can show 
\begin{eqnarray}
  S_{pq}I_{p q r}=I_{q p r} S_{pq}, \quad
  T_{pr} I_{q p r}=I_{q r p} T_{pr}, \quad
  S_{qr} I_{q r p}=I_{r q p} S_{qr}, 
\label{e44}
\end{eqnarray}
which gives 
\begin{eqnarray}
 S_{qr} T_{pr} S_{pq} I_{p q r}=I_{r q p} S_{qr} T_{pr} S_{pq}.
\label{e45}
\end{eqnarray}
From Eq.(\ref{e43}) and Eq.(\ref{e45}), we have 
\begin{eqnarray}
\left[ \left( S_{qr} T_{pr} S_{pq} \right)^{-1}
 T_{pq} S_{pr} T_{qr} , I_{p q r} \right]=0 . 
\label{e46}
\end{eqnarray}

In the same way, using the formula
\begin{eqnarray}
 && X^{-1} (d_p b_q - \omega a_p c_q Z^{-1}) T_{pq}
  =T_{pq} (b_p d_q - c_p a_q Z^{-1}) X^{-1} , 
\label{47}\\
 && Z^{-1} (a_p d_q -c_p b_q X^{-1}/\omega) S_{pq}
  =S_{pq} (d_p a_q - b_p c_q X^{-1}) Z^{-1} , 
\label{48}
\end{eqnarray}
and operate $T_{pq} S_{pr} T_{qr}$ and 
$S_{qr} T_{pr} S_{pq}$ on the quantity $J_{p q r}$
\begin{eqnarray}
 J_{p q r}=-b_p b_q d_r X^{-1}/\omega
 +b_p c_q a_r X^{-1} Z^{-1} -d_p a_q a_r Z^{-1} , 
\label{49}
\end{eqnarray}
we have 
\begin{eqnarray}
  T_{qr}J_{p q r}=J_{p r q} T_{qr}, \quad
  S_{pr}J_{p r q}=J_{r p q} S_{pr}, \quad
  T_{pq}J_{r p q}=J_{r q p} T_{pq}, 
\label{e50}
\end{eqnarray}
and
\begin{eqnarray}
  S_{pq}J_{p q r}=J_{q p r} S_{pq}, \quad
  T_{pr} J_{q p r}=J_{q r p} T_{pr}, \quad
  S_{qr} J_{q r p}=J_{r q p} S_{qr} . 
\label{e51}
\end{eqnarray}
From Eq.(\ref{e50}) and Eq.(\ref{e51}), we have 
\begin{eqnarray}
\left[ \left( S_{qr} T_{pr} S_{pq} \right)^{-1}
 T_{pq} S_{pr} T_{qr} , J_{p q r} \right]=0 .
\label{e52}
\end{eqnarray}
If we notice the relation 
\begin{eqnarray}
  &&\left[ I_{p q r}, J_{p q r} \right]=\frac{\omega -1}{\omega^2}
 \left( a_p b_p b_q d_q b_r d_r Z 
 - \omega a_p d_p a_q d_q a_r b_r X
 - \omega a_p b_p a_q c_q a_r c_r Z^{-1} \right. \nonumber\\
 && \left. + \omega a_p d_q {a_q}^{2} a_r c_r X Z^{-1} 
  +b_p c_p b_q c_q a_r b_r X^{-1} 
 -b_p c_p {b_q}^2 b_r d_r X^{-1} Z/ \omega \right) \ne 0  ,
 \nonumber
\end{eqnarray}
and the cyclic representation is generated by 
 $ \left\{ X, Z \right\}$,  we can conclude from Eq.(\ref{e52}) 
that the operator 
\begin{eqnarray}
 \left( S_{qr} T_{pr} S_{pq} \right)^{-1}
 T_{pq} S_{pr} T_{qr} , \nonumber
\end{eqnarray}
commute with the non-commutative quantities 
$\left\{I_{p q r} , J_{p q r} \right\}$, which means 
\begin{eqnarray}
\left( S_{qr} T_{pr} S_{pq} \right)^{-1}
 T_{pq} S_{pr} T_{qr}=\rho_0 \times {\bf 1}  .
 \label{e53}
\end{eqnarray}
Rewriting the above relation with the normalized 
quantities $\widehat{T}$ and 
$\widehat{S}$, we have  
\begin{eqnarray}
 \widehat{T}_{pq} \widehat{S}_{pr} 
 \widehat{T}_{qr}
 =\rho'_0  \widehat{S}_{qr} \widehat{T}_{pr} 
 \widehat{S}_{pq} .  
\label{e54}
\end{eqnarray}
Taking the determinant of the both side of Eq.(\ref{e54}), we 
have ${\rho'_0}^N=1$, which gives $\rho'_0=\omega^{m}$,
where $m={\rm (integer)}$. If we take the special limit 
$q \rightarrow p$, we have $T_{p p}=1$ and $S_{p p}=1$, 
which gives $\left. \rho'_0 \right|_{q \rightarrow p}=1$, 
but the integer $m$ does not 
change in the limit $ q \rightarrow p$, which gives 
$ \rho'_0=1$ in general. 

In this way, we have proved the star-triangle relation of the 
chiral Potts model in the form 
\begin{eqnarray}
 \widehat{T}_{pq} \widehat{S}_{pr} 
 \widehat{T}_{qr}
 =\widehat{S}_{qr} \widehat{T}_{pr} 
 \widehat{S}_{pq} . 
\label{e55}
\end{eqnarray}


\subsection{Constructive understanding of the star-triangle 
relation}

\indent

We start from the quantity $I_{pqr}$ in the 
general form $I_{pqr}=\alpha_{pqr} X + \beta_{pqr} X Z
  +\gamma_{pqr} Z$ and construct $T_{pq}, S_{pq}$ in such 
a way as it satisfies the relation
\begin{eqnarray}
  T_{qr}I_{p q r}=I_{p r q} T_{qr}, \quad
  S_{pq}I_{p q r}=I_{q p r} S_{pq}.
\label{e56}
\end{eqnarray}
Imposing the $Z_N$ periodicity, we will show that 
the constructed $T_{pq}, S_{pq}$
becomes in the form of Eq.(\ref{e12}) and $I_{pqr}$ becomes 
in the form of Eq.(\ref{e41}).

We first construct $T_{pq}$ from Eq.(\ref{e56}), that is,
\begin{eqnarray}
 T_{qr}(Z) ( \beta_{pqr} Z/ \omega + \alpha_{pqr}) X 
 =( \beta_{p r q} Z/ \omega + \alpha_{p r q}) X T_{qr}(Z),
 \quad \gamma_{pqr}=\gamma_{p r q}.
\label{e57}
\end{eqnarray}
This give the expression in the form 
\begin{eqnarray}
  &&T_{qr}=\sum^{N-1}_{k=0} \widetilde{W}_{qr}(k) Z^k, 
\label{e58}\\
  &&\widetilde{W}_{qr}(k)=\prod^{k}_{l=1} 
  \frac{ \beta_{p r q}- \beta_{pqr} \omega^{l-1}}
       { - \alpha_{p r q}+\alpha_{pqr} \omega^{l}}
   =(p-{\rm independent}).
\label{e59}
\end{eqnarray}

Similarly we construct $S_{pq}$ from Eq.(\ref{e56}), that is,
\begin{eqnarray}
  S_{pq}(X) ( \beta_{pqr} X + \gamma_{pqr}) Z 
 =( \beta_{q p r} X + \gamma_{q p r}) Z S_{pq}(X)  ,
   \quad \alpha_{p q r}=\alpha_{q p r} . 
\label{e60}
\end{eqnarray}
This give the expression of the form 
\begin{eqnarray}
  &&S_{pq}=\sum^{N-1}_{k=0} \overline{W}_{pq}(k) X^k, 
\label{e61}\\
  &&\overline{W}_{pq}(k)=\prod^{k}_{l=1} 
  \frac{- \beta_{p q r}+\beta_{ q p r} \omega^{l-1}}
       { \gamma_{p q r}- \gamma_{q p r} \omega^{l}}
  =(r- {\rm independent}) .
\label{e62}
\end{eqnarray}
In order that $ \widetilde{W}_{qr}$ is independent of $p$
in the right-hand side of Eq.(\ref{e59}) and 
$ \overline{W}_{pq}$ is independent of $r$
in the right-hand side of Eq.(\ref{e62}), we can 
parametrize $\alpha_{pqr}, \beta_{pqr}, \gamma_{pqr}$
in the form 
\begin{eqnarray}
 \alpha_{pqr}=A \alpha^{(1)}_p \alpha^{(1)}_q \alpha^{(2)}_r, 
  \quad
 \beta_{pqr}=B \alpha^{(1)}_p \beta^{(1)}_q \gamma^{(1)}_r, 
 \quad
 &&\gamma_{pqr}=C \gamma^{(2)}_p \gamma^{(1)}_q \gamma^{(1)}_r, 
\label{e63}
\end{eqnarray}
where we used the symmetry relations Eq.(\ref{e57}) and Eq.(\ref{e60}).
Denoting $\alpha^{(1)}_p=a_p, \gamma^{(1)}_p=b_p,
\alpha^{(2)}_p=c_p, \beta^{(1)}_p=d_p, \gamma^{(2)}_p=e_p $,
the periodic condition of $\widetilde{W}_{pq}$ and 
$\overline{W}_{pq}$ give
the conditions
\begin{eqnarray}
 \frac{ {b_p}^N {d_q}^N - {d_p}^N {b_q}^N }
        { {c_p}^N {a_q}^N - {a_p}^N {c_q}^N }=1, \quad
 \frac{ {a_p}^N {d_q}^N - {d_p}^N {a_q}^N }
        { {e_p}^N {b_q}^N - {b_p}^N {e_q}^N }=1, 
\label{e64}
\end{eqnarray}
where we impose $(-A)^N=B^N=(-C)^N$, and we choose 
$A=-1, B=1, C=-1/\omega$.
Up to this level, it is not necessary to take 
$e_p = c_p$, and we have the following expression
\begin{eqnarray}
 && I_{p q r}=-a_p a_q c_r X+a_p d_q b_r X Z
   -e_p b_q b_r Z/ \omega , 
\label{e65}\\
  &&T_{qr}=\sum^{N-1}_{k=0} \widetilde{W}_{qr}(k) Z^k, 
  \quad \widetilde{W}_{qr}(k)=\rho'_1 \prod^{k}_{j=1} 
  \frac{b_q d_r -d_q b_r \omega^{j-1}}
    {c_q a_r -a_q c_r \omega^{j}} ,
\label{e66}\\
  &&S_{pq}=\sum^{N-1}_{k=0} \overline{W}_{pq}(k) X^k, 
  \quad \overline{W}_{pq}(k)=\rho_2 \prod^{k}_{j=1} 
 \frac{\omega a_p d_q -d_p a_q \omega^{j}}
               {e_p b_q -b_p e_q \omega^{j}} .  
\label{e67}
\end{eqnarray}

Similarly, we start from the general form 
$J_{pqr}=\alpha'_{pqr} X^{-1}+ \beta'_{pqr} X^{-1} Z^{-1} 
 +\gamma'_{pqr} Z^{-1}$, which satisfies the properties
\begin{eqnarray}
  T_{qr}J_{p q r}=J_{p r q} T_{qr}, \quad
  S_{pq}J_{p q r}=J_{q p r} S_{pr} , 
\label{e68}
\end{eqnarray}
which gives
\begin{eqnarray}
  &&T_{qr}=\sum^{N-1}_{k=0} \widetilde{W}_{qr}(k) Z^k, 
\label{e69}\\
  &&\widetilde{W}_{qr}(k)=\rho'_1\prod^{k}_{l=1} 
  \frac{ - \omega \alpha'_{pqr}+\alpha'_{p r q} \omega^{l}}
{ \beta'_{pqr}- \beta'_{p r q} \omega^{l}}
          =(p-{\rm independent}),
\label{e70}\\
  &&S_{pq}=\sum^{N-1}_{k=0} \overline{W}_{pq}(k) X^k, 
\label{e71}\\
  &&\overline{W}_{pq}(k)=\rho_2 \prod^{k}_{l=1} 
  \frac{\omega \gamma'_{q p r}- \gamma'_{pqr} \omega^{l}}
       {- \beta'_{pqr}+\beta'_{pqr} \omega^{l}}
  =(r- {\rm independent}) ,
\label{e72}
\end{eqnarray}
with $\gamma'_{pqr}=\gamma'_{p r q}$ and
$\alpha'_{pqr}=\alpha'_{q p r}$.
Then we can parametrize
\begin{eqnarray}
 \alpha'_{pqr}=A' \alpha'^{(1)}_p \alpha'^{(1)}_q \alpha'^{(2)}_r, 
  \quad
 \beta'_{pqr}=B' \alpha'^{(1)}_p \beta'^{(1)}_q \gamma'^{(1)}_r, 
 \quad
 &&\gamma'_{pqr}=C' \gamma'^{(2)}_p \gamma'^{(1)}_q \gamma'^{(1)}_r .
\label{e73}
\end{eqnarray}
We impose that $\tilde{W}_{qr}(k)$ in Eq.(\ref{e65}) and 
Eq.(\ref{e70}) 
gives the same expression and $\overline{W}_{pq}(k)$ in
Eq.(\ref{e67}) and Eq.(\ref{e72}) gives the same 
expression, we obtain $\gamma'^{(1)}_p=a_p, 
\alpha'^{(1)}_p=b_p, \beta'^{(1)}_p=c_p=e_p, 
\alpha'^{(2)}_p=\gamma'^{(2)}_p=d_p$
by choosing $A'=-1/\omega, B'=1, C'=-1$.
In this way, $e_p=c_p$ become necessary to exist 
$J_{pqr}$, we have the desired expression 

\begin{eqnarray}
 && I_{p q r}=-a_p a_q c_r X+a_p d_q b_r X Z
   -c_p b_q b_r Z/ \omega , 
\label{e74}\\
 && J_{p q r}=-b_p b_q d_r X^{-1}/\omega
  +b_p c_q a_r X^{-1} Z^{-1}
   -d_p a_q a_r Z^{-1}  ,
\label{e75}\\
  &&T_{qr}=\sum^{N-1}_{k=0} \widetilde{W}_{qr}(k) Z^k, 
  \quad \widetilde{W}_{qr}(k)=\rho'_1 \prod^{k}_{j=1} 
  \frac{b_q d_r -d_q b_r \omega^{j-1}}
    {c_q a_r -a_q c_r \omega^{j}},
\label{e76}\\
  &&S_{pq}=\sum^{N-1}_{k=0} \overline{W}_{pq}(k) X^k, 
  \quad \overline{W}_{pq}(k)=\rho_2 \prod^{k}_{j=1} 
 \frac{\omega a_p d_q -d_p a_q \omega^{j}}
               {c_p b_q -b_p c_q \omega^{j}},  
\label{e77}
\end{eqnarray}

and the periodic condition becomes 

\begin{eqnarray}
 \frac{ {b_p}^N {d_q}^N - {d_p}^N {b_q}^N }
        { {c_p}^N {a_q}^N - {a_p}^N {c_q}^N }=1, \quad
 \frac{ {a_p}^N {d_q}^N - {d_p}^N {a_q}^N }
        { {c_p}^N {b_q}^N - {b_p}^N {c_q}^N }=1, 
\label{e78}
\end{eqnarray}
   
which is satisfied by the condition
\begin{eqnarray}
  {a_p}^N+k'{b_p}^N=k {d_p}^N, \quad
  k' {a_p}^N+{b_p}^N=k {c_p}^N, 
\label{e79}
\end{eqnarray}
where $\left\{ k, k' \right\}$ are constants,  which are 
not always necessary to satisfy $k^2+{k'}^2=1$ in general.
The rest of the proof of the star-triangle relation 
is the same as that in the previous section.


\subsection{Some generalization}

\indent

From the construction of the previous section, we can 
construct the some generalized model for 
$N=u v \ (u, v : {\rm relatively \ prime \ integers})$. 
For this case, we consider $X'=X^u$, $Z'=Z^u$ and 
these satisfy $Z' X' = \omega' X' Z'$ with 
$\omega' = \omega^{u^2}$. Then we replace 
$X \rightarrow X'$, $Z \rightarrow Z'$, 
$\omega \rightarrow \omega' $ in $T_{qr}$ and $S_{pq}$,
which gives 
\begin{eqnarray}
  &&T_{qr}=\sum^{v-1}_{k=0} \widetilde{W}_{qr}(k) Z^{u k}, 
  \quad \widetilde{W}_{qr}(k)=\rho'_1 \prod^{k}_{j=1} 
  \frac{b_q d_r -d_q b_r \omega^{u^2(j-1)}}
    {c_q a_r -a_q c_r \omega^{u^2 j}},
\label{e80}\\
  &&S_{pq}=\sum^{v-1}_{k=0} \overline{W}_{pq}(k) X^{u k}, 
  \quad \overline{W}_{pq}(k)=\rho_2 \prod^{k}_{j=1} 
 \frac{\omega^{u^2} a_p d_q -d_p a_q \omega^{u^2 j}}
               {c_p b_q -b_p c_q \omega^{u^2 j}}.  
\label{e81}
\end{eqnarray}
If we notice that $u$ and $v$ are relatively prime, 
the periodic condition is
\begin{eqnarray}
  {a_p}^{v}+k'{b_p}^{v}=k {d_p}^{v}, \quad
  k' {a_p}^{v}+{b_p}^{v}=k {c_p}^{v}. 
\label{e82}
\end{eqnarray}
We define the normalized quantity as before 
\begin{eqnarray}
\widehat{T}_{pq}
     =\frac{T_{pq}}{\left|\det(T_{pq})\right|^{1/N}} , 
 \quad 
 \widehat{S}_{pq}
  =\frac{ S_{pq}}{\left|\det(S_{pq})\right|^{1/N}} , 
\label{e83}
\end{eqnarray}
we can show the star-triangle relation of the 
form 
\begin{eqnarray}
  && \widehat{S}_{qr} \widehat{T}_{pr} \widehat{S}_{pq}
  =\widehat{T}_{pq} \widehat{S}_{pr} \widehat{T}_{qr},  
\label{e84}
\end{eqnarray}
for this generalized model. 
In the special case of $ u = 1, v = N $, this reduce the 
original chiral Potts model.


\section{Summary and discussion} 

\indent

We give the simple proof of the star-triangle relation of the 
chiral Potts model. We also give the constructive way to understand 
the star-triangle relation for the chiral Potts model, which may 
give the hint to give the new integrable model. 
We can rewrite the star-triangle relation with the group element 
of the cyclic representation of $su(2)$. Our approach may 
give the hint to give the new two dimensional integrable model
by considering the cyclic reresentation of more general Lie 
algebra. We can prove the 
Yang-Baxter relation of the vertex type chiral Potts model
from the star-triangle relation\cite{B-S}, which suggests that 
the star-triangle relation of the chiral Potts model is more 
fundamental than the Yang-Baxter relation of the vertex type 
chiral Potts model. As the Bazhanov-Baxter model\cite{B-B} is 
constructed from the vertex type chiral Potts model of $sl(n)$, 
we expect that the mathematical structure of Bazhanov-Baxter 
model is understood from the star-triangle relation of the 
chiral Potts model.

\vskip 10mm

\section*{Acknowledgements}

\indent

One of the authors (K. S.) is grateful to the Academic Research Fund
at Tezukayama University for financial support.


\end{document}